\documentclass{ws-p10x7}
\def\Journal#1#2#3#4{{#1} {\bf #2}, #3 (#4)}

\def\NIMA{{\em Nucl. Instrum. Methods} A}
\def\NPB{{\em Nucl. Phys.} B}

\def\PRL{\em Phys. Rev. Lett.}
\def\PRD{{\em Phys. Rev.} D}

\begin{document}

\title{New $R$ values in 2-5 GeV from the BESII at BEPC}

\author{Z. G. Zhao representing the BES Collaboration}

\address{Institute of High Energy Physics, 19 Yuquan Road, Beijing 100039,
China\\E-mail: zhaozg@pony1.ihep.ac.cn}

\twocolumn[\maketitle\abstract{
We report the preliminary $R$ values for all the 85 energy points scanned in
the energy region of 2-5 GeV with the upgraded Beijing Spectrometer (BESII)
at Beijing Electron Positron Collider (BEPC). On average, the uncertainties
of the $R$ values we measured are $\sim 7\%$. The
new $R$ values has a significant impact on the predicted mass of the Higgs
($m_H$) from the global fit to the electroweak data, and will also 
contribute to the interpretation of the E821 $g-2$ experiment.}]

\section{Introduction}
The QED running coupling constant evaluated at the $Z$ pole, 
$\alpha(M^2_{Z})$, and the anomalous 
magnetic moment of the muon, $a_{\mu}=(g-2)/2$, are two fundamental 
quantities to test the Standard Model(SM)~\cite{blondel,zhao}. 
$\alpha(M^2_{Z})$, as 
one of the three input parameters in the global fit to the 
electroweak data, is sensitive to the predicted mass of the Higgs.
Theoretically, $a_{\mu}$ is sensitive to large energy 
scales and very high order radiative corrections~\cite{carey}.
Any deviation between the SM predicted value of anomalous magnetic 
moment of the muon, $a_{\mu}^{SM}$, and that from the experimentally 
measured one, $a_{\mu}^{Exp}$, may hint new physics.  However, the 
uncertainties in both $\alpha(M^2_{Z})$ and $a_{\mu}^{SM}$ are 
dominated 
by the hadronic vacuum polarization, which cannot be reliably 
calculated but are related to $R$ values through dispersion 
relations~\cite{zhao}. Here $R$ is the lowest order cross section for
$e^+e^-\rightarrow\gamma^*\rightarrow \mbox{hadrons}$, which is
defined as
$R=\sigma(e^+e^- \rightarrow \mbox{hadrons})/\sigma(e^+e^-\rightarrow
\mu^+\mu^-)$,
where the denominator is the lowest-order QED cross section,
$\sigma (e^+e^- \rightarrow \mu^+\mu^-) = \sigma^0_{\mu \mu}=
4\pi \alpha^2 / 3s$.
   
Since the uncertainties in $\alpha(M^2_{Z})$ and
$a^{SM}_{\mu}$ are dominated by the errors of the 
values of $R$ in the cm energy range below 5 GeV~\cite{zhao}, it is 
crucial to significantly reduce the uncertainties in the $R$ 
values measured about 20 years ago with a precision of about 
15-20\% in the energy region of 2-5 GeV~\cite{blondel,carey}.  

\section{$R$ scan with BESII at BEPC}

Following the first $R$ scan with 6 energy points in 2.6-5 GeV range 
done in 1998~\cite{besr_1}, the BES collaboration did a finer 
$R$ scan with 85 energy points in the energy region of 2-4.8 GeV.
To understand the beam associated background,
separated beam running was done at 24 energy points and single
beam running for both $e^-$ and $e^+$ was done at 7 energy
points distributed over the whole scanned energy region. Special runs
were taken at the $J/\psi$ to determine the trigger efficiency. 

The scan was done with BESII, a conventional collider detector 
which has been described in detail in ref. 5. 

\section{Data Analysis} 

The $R$ values from the BESII scan data are measured by observing the
final hadronic events inclusively, i.e. the value of $R$ is determined
from the number of observed hadronic events ($N^{obs}_{had}$) by the
relation
\begin{equation}
R=\frac{ N^{obs}_{had} - N_{bg} - \sum_{l}N_{ll} - N_{\gamma\gamma} }
{ \sigma^0_{\mu\mu} \cdot L \cdot \epsilon_{had} 
\cdot (1+\delta)},
\end{equation}
where $N_{bg}$ is the number of beam associated background events;
$\sum_{l}N_{ll},~(l=e,\mu,\tau)$ and $N_{\gamma\gamma}$ are the numbers
of misidentified lepton-pairs from one-photon and two-photon processes
events respectively; $L$ is the integrated luminosity; $\delta$ is
the radiative correction; $\epsilon_{had}$ is the detection efficiency 
for hadronic events. 

The hadronic event selection identifies one photon
multi-hadron production from all other possible contamination
mechanisms. Cosmic rays, lepton pair production, two-photon 
process and beam associated processes are the backgrounds involved 
in our measurement. Clear Bhabha events are first 
rejected.  Then the hadronic events are selected based on charged track information. Cuts on fiducial, 
vertex, track fit 
quality, maximum and minimum energy deposition, momentum and 
time-of-flight are applied to select hadronic events. Special 
attention is paid to two-prong events. Additional 
cuts are utilized to further reject cosmic ray, Bhabha and beam 
associated backgrounds~\cite{besr_1}.     

The cosmic rays and part of the lepton pair production events are
directly removed by the event selection. The remaining background from
lepton pair production and two-photon processes is then subtracted out
statistically according to a Monte Carlo simulation.

Most of the beam associated background events are rejected 
by a vertex cut.  
The same hadronic event selection criteria 
are applied to the separated-beam data, and the number of 
separated-beam events, $N_{sep}$, surviving these criteria are 
obtained.  The number of the beam
associated background events, $N_{bg}$, in the corresponding hadronic
event sample is given by $N_{bg}=f \times N_{sep}$, where $f$ is
the ratio of the product of the pressure at the
collision region times the integrated beam currents for colliding beam
runs and that for the separated beam runs.

The beam associated background can also be subtracted by fitting the
event vertex along the beam direction with a Gaussian for real hadronic
events and a polynomial of degree two for the background. 
The difference between $R$ values obtained using these two methods to
subtract the beam associated background is about ($0.3 \sim 2.3)\%$,
depending on the energy. This difference was taken into account in the
systematic uncertainty.

The integrated luminosity is determined by large-angle Bhabha events
using only the energies deposited in BSC.

A cross check using only $dE/dx$ information from the MDC
to identify electrons was generally consistent with the BSC 
measurement; the difference was taken into account in the overall 
systematic error of 1.5-2.6\%.

A special effort has been made by the Lund group and BES collaboration to
develop the formalism using the basic Lund Model Area Law directly for
a Monte Carlo simulation, which removes the extreme high energy 
approximation in string fragmentation in
JETSET~\cite{bo}. The final state simulation in LUARLW is exclusive 
as opposed to JETSET's inclusive method, and LUARLW has only one free 
parameter in the fragmentation function.  Above 3.77 GeV, the 
production of charmed mesons
are included in the generator based on Eichten Model~\cite{eichiten,chenjc}.

The parameters in LUARLW are tuned with $R$ scan data to
reproduce distributions of kinematic variables such as multiplicity,
sphericity, angular and momentum distributions, etc. 

We find that the same set of parameters can be applied to the energy
region below open charm, and another set of parameters can be used for
the energies above it. Parameters are also tuned point by point for the
continuum and we find that the detection efficiencies determined are
consistent within 2\%, which is taken into account in the systematic
errors.

Radiative corrections determined using four different
schemes~\cite{besr_1} 
agreed with each other to within 1\%
below charm threshold.  Above charm threshold, where resonances are
important, the agreement is within 1-3\%.  
For the measurements reported here, we
use the formalism of Ref. 9 and include the differences with
the other schemes in the systematic error of less than 4\%.

\begin{table}
\caption{Error Sources for $E_{cm}$=3.0 GeV. Adding the systematic 
and statistic errors in quadrature gives a total error of 5.8\%.}
\begin{center}
\begin{tabular}{cccccc} \hline
Source & $N_{had}$ & $L$ & $\epsilon_{had}$ & $1+\delta$ & Stat.\\
Err.(\%) & 3.3 & 2.3 & 3.0 & 1.3 & 2.5\\ \hline
\end{tabular}
\end{center}
\end{table}

The errors from different sources are listed in table 1. 

To further improve the measurement of $R$ values at BEPC, one needs
better performance from the detector and a better handle on the
uncertainty arising from the hadronic event generator, as well as
higher machine luminosity, particularly for the energies below
3.0 GeV.

\begin{figure}
\centerline{\psfig{figure=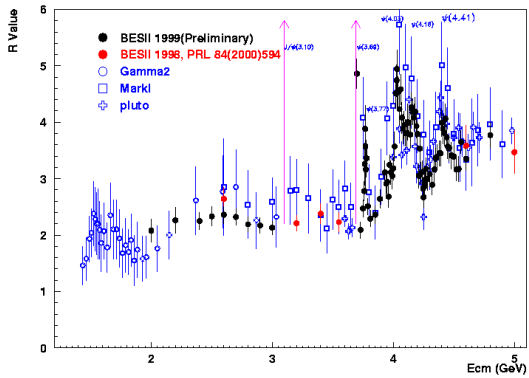,width=60mm,angle=-90}}
\caption{$R$ values below 5 GeV. \label{fig:besr}}
\end{figure}

The preliminary $R$ values obtained at all 85 energy are 
graphically displayed in Fig.~\ref{fig:besr}, together with the 
6 energy points measured in the first scan and those measured by MarkI,
$\gamma\gamma 2$ and Pluto about twenty years ago. 
The preliminary $R$ values from BESII have an average uncertainty of
about 7\% 
and are slightly lower than that from the previously measurements. The
two to three factor improvement in precision of the $R$ values in 2-5 GeV 
has a significant impact on the global fit to the electroweak data for the
determination of $m_H$. The preliminary fit results show that the
predicted $m_H$ is significant increased with the preferred
value lying just above the LEP2 excluded region, and the new $\chi^2$ profile
of the fit accommodates the LEP2 bound on the mass more
comfortably~\cite{bolek,martin}. 
On the other hand, BESII $R$ values can also greatly contribute to the
interpretation of the E821 $g-2$ measurement~\cite{carey}.  

We would like to thank the staff of the BEPC accelerator and IHEP Computing
Center for their efforts.  
We also wish to acknowledge useful discussions
with B. Andersson, H. Burkhardt, M. Davier, B. Pietrzyk,
T. Sj\"{o}strand,  A. D. Martin, M. L. Swartz.
We especially thank M. Tigner for major contributions not only to
BES but also to the operation of the BEPC.


\vspace*{0.2cm}

\begin{thebibliography}{99}
\bibitem{blondel}A. Blondel, Proc. of the 28th Int. Conf. on High Energy
Physics, Warsaw, Poland, July 1996. 
\bibitem{zhao}Z. G. Zhao, Proc. of LP99, SLAC, USA, July 1999. 
\bibitem{carey} Robert Carey, "New Results from g-2 Experiment", talk given 
at ICHEP2000, Osaka, Japan, July 2000.
\bibitem{besr_1} J. Z. Bai, et. al. the BES collaboration, 
\Journal{\PRL}{84}{594}{2000}.   
\bibitem{bes}J.Z. Bai {\it et al.}, (BES Collab.),
\Journal{\NIMA}{344}{319}{1994};
J.Z. Bai {\it et al.}, (BES Collab.), "The Upgraded Beijing Spectrometer",
accepted by NIM.
\bibitem{bo} B. Andersson and Haiming Hu, "Few-body States in Lund String
Fragmentation Model" hep-ph/9910285.
\bibitem{eichiten}E. Eichten, et.,al., \Journal{\PRD}{21}{203}{1980}.
\bibitem{chenjc}J.C. Chen, et, al., \Journal{\PRD}{62}{034003}{2000}.
\bibitem{9}G. Bonneau and F. Martin, \Journal{\NPB}{27}{387}{1971}.
\bibitem{bolek} B. Pietrzyk, "The Global Fit to the Electroweak Data", 
talk given at ICHEP2000, Osaka, Japan, July 2000.
\bibitem{martin} A. Martin, et. al., hep-ph/0008078. 
\end{thebibliography}
\end{document}